\documentclass[10pt]{article}
\usepackage{latexsym,graphicx}
\newcommand{\be}{\begin{equation}}
\newcommand{\ee}{\end{equation}}
\def\n{\noindent}
\catcode `\@=11
\catcode `\@=12
\begin{document}
\begin{center}
\large{\bf {A New Class of Inhomogeneous String Cosmological Models in General 
Relativity}} \\
\vspace{10mm}
\normalsize{Anirudh Pradhan \footnote{Corresponding author}, Anil Kumar Yadav $^2$, 
R. P. Singh $^3$ and Vipin Kumar Singh $^4$}\\
\vspace{5mm}
\normalsize{$^{1}$ Department of Mathematics, Hindu Post-graduate College, 
Zamania-232 331, Ghazipur, India \\
E-mail : pradhan@iucaa.ernet.in} \\
\vspace{5mm}
\normalsize{$^{2}$Department of Physics, K. N. Govt. Post-graduate College, 
Sant Ravidas Nagar (Gyanpur), Bhadohi - 221 304, India \\
E-mail : abanilyadav@yahoo.co.in}\\ 
\vspace{5mm}
\normalsize{$^{3,4}$ Department of Mathematics, T. D. Post-graduate College, 
Jaunpur-222 002, India} \\
\end{center}
\vspace{10mm}
\begin{abstract} 
A new class of solutions of Einstein field equations has been investigated for 
inhomogeneous cylindrically symmetric space-time with string source. To get the 
deterministic solution, it has been assumed that the expansion ($\theta$) in the 
model is proportional to the eigen value $\sigma^{1}~~_{1}$ of the shear tensor 
$\sigma^{i}~~_{j}$. Certain physical and geometric properties of the models are 
also discussed.   
\end{abstract}
\smallskip
\n Keywords : String, Inhomogeneous universe, Cylindrical symmetry\\
\n PACS number: 98.80.Cq, 04.20.-q 
\section{Introduction}
In recent years, there has been considerable interest in string cosmology because 
cosmic strings play an important role in the study of the early universe. These 
strings arise during the phase transition after the big bang explosion as the 
temperature goes down below some critical temperature as predicted by grand unified 
theories (Zel'dovich et al., 1975; Kibble, 1976, 1980; Everett, 1981; Vilenkin, 1981). 
Moreover, the investigation of cosmic strings and their physical processes near such 
strings has received wide attention because it is believed that cosmic strings give 
rise to density perturbations which lead to formation of galaxies (Zel'dovich, 1980; 
Vilenkin, 1981). These cosmic strings have stress energy and couple to the 
gravitational field. Therefore, it is interesting to study the gravitational effect 
which arises from strings by using Einstein's equations. \\

The general treatment of strings was initiated by Letelier (1979, 1983) and 
Stachel (1980). Letelier (1979) obtained the general solution of Einstein's 
field equations for a cloud of strings with spherical, plane and a particular 
case of cylindrical symmetry. Letelier (1983) also obtained massive string 
cosmological models in Bianchi type-I and Kantowski-Sachs space-times. Benerjee 
et al. (1990) have investigated an axially symmetric Bianchi type I string dust 
cosmological model in presence and absence of magnetic field using a supplementary 
condition $\alpha = a \beta$ between metric potential where $\alpha = \alpha(t)$ and 
$\beta = \beta(t)$ and $a$ is constant. Exact solutions of string cosmology for 
Bianchi type-II, $-VI_{0}$, -VIII and -IX space-times have been studied by Krori et al. 
(1990) and Wang (2003). Wang (2004, 2005, 2006) has investigated bulk viscous string 
cosmological models in different space-times. Bali et al. (2001, 2003, 2005, 2006, 2007)
have obtained Bianchi type-I, -III, -V and type-IX string cosmological models in general 
relativity. The string cosmological models with a magnetic field are discussed by 
Chakraborty (1991), Tikekar and Patel (1992, 1994), Patel and Maharaj (1996). Ram 
and Singh (1995) obtained some new exact solution of string cosmology with and without a 
source free magnetic field for Bianchi type I space-time in the different basic form 
considered by Carminati and McIntosh (1980). Singh and Singh (1999) investigated string 
cosmological models with magnetic field in the context of space-time with $G_{3}$ 
symmetry. Singh (1995) has studied string cosmology with electromagnetic fields in 
Bianchi type-II, -VIII and -IX space-times. Lidsey, Wands and Copeland (2000) have 
reviewed aspects of super string cosmology with the emphasis on the cosmological 
implications of duality symmetries in the theory. Yavuz et al. (2005) have examined 
charged strange quark matter attached to the string cloud in the spherical symmetric 
space-time admitting one-parameter group of conformal motion. Recently Kaluza-Klein 
cosmological solutions are obtained by Yilmaz (2006) for quark matter attached to the 
string cloud in the context of general relativity. \\   

Cylindrically symmetric space-time play an important role in the study of the universe 
on a scale in which anisotropy and inhomogeneity are not ignored. Inhomogeneous 
cylindrically symmetric cosmological models have significant contribution in 
understanding some essential features of the universe such as the formation of 
galaxies during the early stages of their evolution. Bali and Tyagi (1989) and 
Pradhan et al. (2001, 2006) have investigated cylindrically symmetric inhomogeneous 
cosmological models in presence of electromagnetic field. Barrow and Kunze (1997, 1998) 
found a wide class of exact cylindrically symmetric flat and open inhomogeneous string 
universes. In their solutions all physical quantities depend on at most one space 
coordinate and the time. The case of cylindrical symmetry is natural because of the 
mathematical simplicity of the field equations whenever there exists a direction in 
which the pressure equal to energy density. \\

Recently Baysal et al. (2001), Kilinc and Yavuz (1996) have investigated 
some string cosmological models in cylindrically symmetric inhomogeneous universe. In 
this paper, we have revisited their solutions and obtained a new class of solutions. 
Here, we extend our understanding of inhomogeneous string cosmologies by investigating 
the simple models of non-linear cylindrically symmetric inhomogeneities outlined above. 
This paper is organized as follows: The metric and field equations are presented in 
Section $2$. In Section $3$, we deal with the solution of the field equations in two 
different cases. Finally, the results are discussed in Section $4$. The solutions 
obtained in this paper are new and different from the other author's solutions.  
\section{The Metric and Field  Equations}
We consider the Bianchi Type I metric in the form 
\begin{equation}
\label{eq1}
ds^{2} = A^{2}(dx^{2} - dt^{2}) + B^{2} dy^{2} + C^{2} dz^{2},
\end{equation}
where $A$, $B$ and $C$ are functions of $x$ and $t$.
The Einstein's field equations for a cloud of strings read as (Letelier, 1983) 
\begin{equation}
\label{eq2}
G^{j}_{i} \equiv  R^{j}_{i} - \frac{1}{2} R g^{j}_{i} = -(\rho u_{i}u^{j} - 
\lambda x_{i}x^{j}),
\end{equation}
where $u_{i}$ and $x_{i}$ satisfy conditions
\begin{equation}
\label{eq3}
u^{i} u_{i} = - x^{i} x_{i} = -1,
\end{equation}
and
\begin{equation}
\label{eq4}
u^{i} x_{i} = 0.
\end{equation}
Here, $\rho$ is the rest energy of the cloud of strings with massive particles 
attached to them. $\rho = \rho_{p} + \lambda$, $\rho_{p}$ being the rest energy 
density of particles attached to the strings and  $\lambda$ the density of tension 
that characterizes the strings. The unit space-like vector  $x^{i}$ represents 
the string direction in the cloud, i.e. the direction of anisotropy and the 
unit time-like vector  $u^{i}$ describes the four-velocity vector of the matter 
satisfying the following conditions 
\begin{equation}
\label{eq5}
g_{ij} u^{i} u^{j} = -1.
\end{equation}
In the present scenario, the comoving coordinates are taken as 
\begin{equation}
\label{eq6}
u^{i} = \left(0, 0, 0, \frac{1}{A}\right) 
\end{equation}
and choose $x^{i}$ parallel to x-axis so that
\begin{equation}
\label{eq7}
x^{i} = \left(\frac{1}{A}, 0, 0, 0 \right). 
\end{equation}
The Einstein's field equations (\ref{eq2}) for the line-element (\ref{eq1}) 
lead to the following system of equations:  
\[
G^{1}_{1} \equiv \frac{B_{44}}{B} + \frac{C_{44}}{C} - \frac{A_{4}}{A}
\left(\frac{B_{4}}{B} + \frac{C_{4}}{C}\right) - \frac{A_{1}}{A}
\left(\frac{B_{1}}{B} + \frac{C_{1}}{C}\right) -\frac{B_{1}C_{1}}{BC} + 
\frac{B_{4} C_{4}}{B C} 
\]
\begin{equation}
\label{eq8}
= \lambda A^{2},
\end{equation}
\begin{equation}
\label{eq9}
G^{2}_{2} \equiv \left(\frac{A_{4}}{A}\right)_{4} - \left(\frac{A_{1}}
{A}\right)_{1} + \frac{C_{44}}{C} - \frac{C_{11}}{ C} = 0,
\end{equation}
\begin{equation}
\label{eq10}
G^{3}_{3} \equiv \left(\frac{A_{4}}{A}\right)_{4} - \left(\frac{A_{1}}{A}\right)_{1} 
+ \frac{B_{44}}{B} - \frac{B_{11}}{B} =  0,
\end{equation}
\[
G^{4}_{4} \equiv - \frac{B_{11}}{B} - \frac{C_{11}}{C} + \frac{A_{1}}{A}
\left(\frac{B_{1}}{B} + \frac{C_{1}}{C}\right) + \frac{A_{4}}{A}\left(\frac{B_{4}}{B} 
+ \frac{C_{4}}{C}\right) - \frac{B_{1}C_{1}}{BC}  + \frac{B_{4} C_{4}}{B C} 
\]
\begin{equation}
\label{eq11}
= \rho A^{2},
\end{equation}
\begin{equation}
\label{eq12}
G^{1}_{4} \equiv \frac{B_{14}}{B} + \frac{C_{14}}{C} - \frac{A_{4}}{A}\left(\frac{B_{1}}{B}
 + \frac{C_{1}}{C}\right) - \frac{A_{1}}{A}\left(\frac{B_{4}}{B} + \frac{C_{4}}{C}\right) = 0,
\end{equation}
where the sub indices $1$ and $4$ in A, B, C and elsewhere denote differentiation
with respect to $x$ and $t$, respectively.

The velocity field $u^{i}$ is irrotational. The scalar expansion $\theta$, shear scalar 
$\sigma^{2}$, acceleration vector $\dot{u}_{i}$ and proper volume $V^{3}$ are respectively 
found to have the following expressions:
\begin{equation}
\label{eq13}
\theta = u^{i}_{;i} = \frac{1}{A}\left(\frac{A_{4}}{A} + \frac{B_{4}}{B} + \frac{C_{4}}{C}
\right),
\end{equation}
\begin{equation}
\label{eq14}
\sigma^{2} = \frac{1}{2} \sigma_{ij} \sigma^{ij} = \frac{1}{3}\theta^{2} - \frac{1}{A^{2}}
\left(\frac{A_{4}B_{4}}{AB} + \frac{B_{4}C_{4}}{BC} + \frac{C_{4}A_{4}}{CA}\right),
\end{equation}
\begin{equation}
\label{eq15}
\dot{u}_{i} = u_{i;j}u^{j} = \left(\frac{A_{1}}{A}, 0, 0, 0\right), 
\end{equation}
\begin{equation}
\label{eq16}
V^{3} = \sqrt{-g} = A^{2} B C,
\end{equation}
where $g$ is the determinant of the metric (\ref{eq1}). Using the field equations and 
the relations (\ref{eq13}) and (\ref{eq14}) one obtains the Raychaudhuri's equation as
\begin{equation}
\label{eq17}
\dot{\theta} = \dot{u}^{i}_{;i} - \frac{1}{3}\theta^{2} - 2 \sigma^{2} - \frac{1}{2} 
\rho_{p},
\end{equation}
where dot denotes differentiation with respect to $t$ and
\begin{equation}
\label{eq18}
R_{ij}u^{i}u^{j} = \frac{1}{2}\rho_{p}.
\end{equation}
 
With the help of equations (\ref{eq1}) - (\ref{eq7}), the Bianchi identity 
$\left(T^{ij}_{;j}\right)$ reduced to two equations:
\begin{equation}
\label{eq19}
\rho_{4} - \frac{A_{4}}{A}\lambda + \left(\frac{A_{4}}{A} + \frac{B_{4}}{B} + 
\frac{C_{4}}{C}\right)\rho = 0
\end{equation}
and
\begin{equation}
\label{eq20}
\lambda_{1} - \frac{A_{1}}{A}\rho + \left(\frac{A_{1}}{A} + \frac{B_{1}}{B} + 
\frac{C_{1}}{C}\right)\lambda = 0.
\end{equation}
Thus due to all the three (strong, weak and dominant) energy conditions, one finds 
$\rho \geq 0$ and $\rho_{p} \geq 0$, together with the fact that the sign of $\lambda$ 
is unrestricted, it may take values positive, negative or zero as well.  
\section{Solutions of the Field  Equations}

As in the case of general-relativistic cosmologies, the introduction of inhomogeneities 
into the string cosmological equations produces a considerable increase in mathematical 
difficulty: non-linear partial differential equations must now be solved. In practice, 
this means that we must proceed either by means of approximations which render the non-
linearities tractable, or we must introduce particular symmetries into the metric of the 
space-time in order to reduce the number of degrees of freedom which the inhomogeneities 
can exploit. \\

Here to get a determinate solution, let us assume that expansion ($\theta$) 
in the model is proportional to the eigen value $\sigma^{1}~~_{1}$ of  the shear tensor 
$\sigma^{i}~~_{j}$. This condition leads to
\begin{equation}
\label{eq21}
A = (BC)^{n},
\end{equation}
where $n$ is a constant. Equations (\ref{eq9}) and (\ref{eq10}) lead to
\begin{equation}
\label{eq22}
\frac{B_{44}}{B} - \frac{B_{11}}{B} = \frac{C_{44}}{C} - \frac{C_{11}}{C}.
\end{equation}
Using (\ref{eq21}) in (\ref{eq12}), yields 
\begin{equation}
\label{eq23}
\frac{B_{41}}{B} + \frac{C_{41}}{C} - 2n \left(\frac{B_{4}}{B} + 
\frac{C_{4}}{C}\right)\left(\frac{B_{1}}{B} + \frac{C_{1}}{C}\right) = 0.
\end{equation}
To find out deterministic solutions, we consider the following three cases: 
$$ (i) B = f(x)g(t) ~ ~ \mbox{and} ~ ~ C = h(x) k(t),$$
$$ (ii) B = f(x)g(t) ~ ~ \mbox{and} ~ ~ C = f(x) k(t),$$
$$ (iii) B = f(x)g(t) ~ ~ \mbox{and} ~ ~ C = h(x) g(t).$$
The case (iii) is discussed by Kilinc and Yavuz (1996). We obtain a new class 
of solutions for other two cases (i) and (ii) and discuss their consequences 
separately below in this paper. 
\subsection{Case(i):} $B = f(x)g(t)$ and $C = h(x)k(t)$ \\

In this case equation (\ref{eq23}) reduces to
\begin{equation}
\label{eq24}
\frac{f_{1}/f}{h_{1}/h} = - \frac{(2n - 1)(k_{4}/k) + 2n(g_{4}/g)}{(2n - 1)(g_{4}/g) + 
2n(k_{4}/k)} = K \mbox{(constant)},
\end{equation}
which leads to
\begin{equation}
\label{eq25}
\frac{f_{1}}{f} = K\frac{h_{1}}{h}
\end{equation}
and
\begin{equation}
\label{eq26}
\frac{k_{4}/k}{g_{4}/g} = \frac{K - 2nK - 2n}{2nK + 2n - 1} = a \mbox{(constant)}.
\end{equation}
From Eqs. (\ref{eq25}) and (\ref{eq26}), we obtain
\begin{equation}
\label{eq27}
f = \alpha h^{K}
\end{equation}
and
\begin{equation}
\label{eq28}
k = \delta g^{a},
\end{equation}
where $\alpha$ and $\delta$ are integrating constants. Eq. (\ref{eq22}) reduces to  
\begin{equation}
\label{eq29}
\frac{g_{44}}{g} - \frac{k_{44}}{k} = \frac{f_{11}}{f} - \frac{h_{11}}{h} = N,
\end{equation}
where $N$ is a constant. Using the functional values of B and C in (\ref{eq22}), 
we obtain 
\begin{equation}
\label{eq30}
g g_{44} + a g_{4}^{2} = - \frac{p^{2}}{(1+ a)}g^{2},
\end{equation}
which leads to 
\begin{equation}
\label{eq31}
g = \left(b_{1} e^{pt} + b_{2} e^{-pt}\right)^{\frac{1}{a+ 1}},
\end{equation}
where
$$ p = \sqrt{\frac{N(a + 1)}{(1 - a)}}$$
and $b_{1}$, $b_{2}$ are constants of integration. Thus from Eq. (\ref{eq28}) we get
\begin{equation}
\label{eq32}
k = \delta \left(b_{1} e^{pt} + b_{2} e^{-pt}\right)^{\frac{a}{a + 1}}.
\end{equation}
From Eqs. (\ref{eq25}) and (\ref{eq29}), we obtain 
\begin{equation}
\label{eq33}
h h_{11} + K h_{1}^{2} = \frac{q^{2}}{(K + 1)}h^{2},
\end{equation}
which leads to
\begin{equation}
\label{eq34}
h = \left(c_{1} e^{qx} + c_{2} e^{-qx}\right)^{\frac{1}{K + 1}},
\end{equation}
where 
$$ q = \sqrt{\frac{N(K + 1)}{(K - 1)}}$$
and $c_{1}$, $c_{2}$ are constants of integration. Hence from Eq. (\ref{eq27}) we have 
\begin{equation}
\label{eq35}
f = \alpha\left(c_{1} e^{qx} + c_{2} e^{-qx}\right)^{\frac{K}{K + 1}}.
\end{equation}
It is worth mentioned here that equations (\ref{eq30}) and (\ref{eq33}) are 
fundamental basic differential equations for which we have reported new 
solutions given by equations (\ref{eq31}) and (\ref{eq34}). \\
Thus, we obtain 
\begin{equation}
\label{eq36}
B = fg =\alpha\left(c_{1} e^{qx} + c_{2} e^{-qx}\right)^{\frac{K}{K + 1}}\left(b_{1} e^{pt} 
+ b_{2} e^{-pt}\right)^{\frac{1}{a+ 1}},
\end{equation}
\begin{equation}
\label{eq37}
C = hk = \delta \left(c_{1} e^{qx} + c_{2} e^{-qx}\right)^{\frac{1}{K + 1}}\left(b_{1} e^{pt} + 
b_{2} e^{-pt}\right)^{\frac{a}{a + 1}}. 
\end{equation}
Therefore
\begin{equation}
\label{eq38}
A = (BC)^{n} = (\alpha \delta)^{n} \left(c_{1} e^{qx} + c_{2} e^{-qx}\right)^{n}\left(b_{1} 
e^{pt} + b_{2} e^{-pt}\right)^{n},
\end{equation}

Hence the metric (\ref{eq1}) takes the form
\[
ds^{2}= (\alpha \delta)^{2n} \left(c_{1} e^{qx} + c_{2} e^{-qx}\right)^{2n}\left(b_{1} e^{pt} + 
b_{2} e^{-pt}\right)^{2n} (dx^{2} - dt^{2}) + 
\]
\[
\left(c_{1} e^{qx} + c_{2} e^{-qx}\right)^{\frac{2}{K + 1}}\left(b_{1} e^{pt} + 
b_{2} e^{-pt}\right)^{\frac{2}{a+ 1}} \Bigl[\alpha^{2}\left(c_{1} e^{qx} + c_{2} 
e^{-qx}\right)^{\frac{2(K - 1)}{K + 1}}dy^{2}
\]
\begin{equation}
\label{eq39}
+ \delta^{2}\left(b_{1} e^{pt} + b_{2} e^{-pt}\right)^{\frac{2(a - 1)}{(a + 1)}}dz^{2}\Bigr].
\end{equation}
The energy density $(\rho)$, the string tension density $(\lambda)$ and the particle 
density $(\rho_{p})$ for the model (\ref{eq46}) are given by 
\[
\rho = \frac{1}{(\alpha \delta)^{2n} \phi(x)^{2n}\psi(t)^{2n}}\Biggl[-\frac{4q^{2}c_{1}c_{2}}
{\phi(x)^{2}} - \frac{q^{2}\eta(x)^{2}}{\phi(x)^{2}}\left\{\frac{K^{2} + K + 1}{(K + 1)^{2}} 
- n \right\} 
\]
\begin{equation}
\label{eq40}
+ \frac{p^{2}\xi(t)^{2}}{\psi(t)^{2}}\left\{\frac{a}{(a + 1)^{2}} + n \right\}\Biggr], 
\end{equation}
\[
\lambda = \frac{1}{(\alpha \delta)^{2n} \phi(x)^{2n}\psi(t)^{2n}}\Biggl[\frac{4p^{2}b_{1}b_{2}}
{\psi(t)^{2}} + \frac{p^{2}\xi(t)^{2}}{{\psi(t)}^{2}}\left\{\frac{a^{2} + a + 1}
{(a + 1)^{2}} - n \right\}
\]
\begin{equation}
\label{eq41}
 + \frac{q^{2}\eta(x)^{2}}{\phi(x)^{2}}\left\{\frac{K}{(K + 1)^{2}} + n \right\}\Biggr], 
\end{equation}
\begin{equation}
\label{eq42}
\rho_{p} = \frac{1}{(\alpha \delta)^{2n} \phi(x)^{2n}\psi(t)^{2n}}\Biggl[-\frac{4q^{2}
c_{1}c_{2}}{\phi(x)^{2}} - \frac{4p^{2}b_{1}b_{2}}{\psi(t)^{2}} - \frac{q^{2}\eta(x)^{2}}
{\phi(x)^{2}}+ \frac{p^{2}\xi(t)^{2}}{\psi(t)^{2}}\Biggr],
\end{equation}
where
$$ \phi(x) = c_{1}e^{qx} + c_{2}e^{-qx},$$
$$ \psi(t) = b_{1}e^{pt} + b_{2}e^{-pt},$$
$$ \eta(x) = c_{1}e^{qx} - c_{2}e^{-qx},$$
\begin{equation}
\label{eq43}
\xi(t) = b_{1}e^{pt} - b_{2}e^{-pt}.
\end{equation}
The scalar of expansion $(\theta)$, shear tensor $(\sigma)$, the acceleration vector 
$(\dot{u}_{i})$ and the proper volume $(V^{3})$ for the model (\ref{eq46}) are obtained as 
\begin{equation}
\label{eq44}
\theta = \frac{p (n + 1)\xi(t)}{(\alpha \delta)^{n} \phi(x)^{n}\psi(t)^{n + 1}}, 
\end{equation}
\begin{equation}
\label{eq45}
\sigma^{2} =  \frac{p^{2}\xi(t)^{2}}{(\alpha \delta)^{2n}\phi(x)^{2n}\psi(t)^{2n + 2}}
\Biggl[\frac{1}{3}(n + 1)^{2} -\left(n + \frac{a}{(a + 1)^{2}}\right)\Biggr], 
\end{equation}
\begin{equation}
\label{eq46}
\dot{u}_{i} = \left(\frac{n q \eta(x)}{\phi(x)}, 0, 0, 0\right),
\end{equation}
\begin{equation}
\label{eq47}
V^{3} = \sqrt{-g} = (\alpha \delta)^{2n + 1}\phi(x)^{2n + 1} \psi(t)^{2n + 1}.
\end{equation}
From equations (\ref{eq44}) and (\ref{eq45}), we obtain
\begin{equation}
\label{eq48}
\frac{\sigma^{2}}{\theta^{2}} = \frac{1}{3} - \frac{n(a + 1)^{2} + a}{(a + 1)^{2}
(n + 1)^{2}} = \mbox{(constant)}.
\end{equation}
\subsection{Case(ii):} $B = f(x)g(t)$ and $C = f(x)k(t)$ \\

In this case equation (\ref{eq23}) reduces to
\begin{equation}
\label{eq49}
(4n -1)\frac{f_{1}}{f}\left(\frac{g_{4}}{g} + \frac{k_{4}}{k}\right) = 0.
\end{equation}
The equation (\ref{eq49}) leads to three cases:
\[
(a) ~ ~ n = \frac{1}{4},
\]
\[
(b) ~ ~ \frac{f_{1}}{f} = 0,
\]
\[
(c) ~~ \frac{g_{4}}{g} + \frac{k_{4}}{k} = 0.
\]
The case (a) reduces the number of equation to four but, with five unknowns which 
requires additional assumption for a viable solution. In the case (b), the model 
turns to be a particular case to the Bianchi type-I model. Therefore we consider 
the case (c) only.\\

Using condition (c) in equation (\ref{eq22}) leads to 
\begin{equation}
\label{eq50}
\frac{g_{44}}{g} = \frac{k_{44}}{k}.
\end{equation}
By using condition (c) in (\ref{eq50}), we get
\begin{equation}
\label{eq51}
g = e^{\ell t}, ~ ~ ~ k = e^{-\ell t},
\end{equation}
where $\ell$ is constant. \\

From equations (\ref{eq9}) or (\ref{eq10}) and (\ref{eq51}), we have
\begin{equation}
\label{eq52}
ff_{11} - \frac{2n}{2n + 1}f^{2}_{1} - \frac{\ell^{2}}{2n + 1}f^{2} = 0.
\end{equation}
Solving (\ref{eq52}), we obtain
\begin{equation}
\label{eq53}
f = \left(d_{1}e^{rx} + d_{2}e^{-rx}\right)^{2n + 1} ,
\end{equation}
where $d_{1}$ and $d_{2}$ are constants of integration and 
$$ r = \frac{\ell}{2n + 1}.$$
 
It is important to mention here that (\ref{eq52}) is the basic equation for which 
new solution is obtained as given by (\ref{eq53}). \\

Hence, we obtain
\begin{equation}
\label{eq54}
B = fg = \left(d_{1}e^{rx} + d_{2}e^{-rx}\right)^{2n + 1} e^{\ell t}
\end{equation}
and
\begin{equation}
\label{eq55}
C = fk = \left(d_{1}e^{rx} + d_{2}e^{-rx}\right)^{2n + 1} e^{-\ell t}. 
\end{equation}
Therefore
\begin{equation}
\label{eq56}
A = (BC)^{n} = \left(d_{1}e^{rx} + d_{2}e^{-rx}\right)^{2n(2n + 1)}.
\end{equation}
In this case the metric (\ref{eq1}) reduces to the form 
\begin{equation}
\label{eq57}
ds^{2} = (d_{1}e^{rx} + d_{2}e^{-rx})^{4n(2n + 1)}(dx^{2} - dt^{2}) + (d_{1}e^{rx} + 
d_{2}e^{-rx})^{2}(e^{2 \ell t}dy^{2} + e^{-2 \ell t}dz^{2}).
\end{equation}

In this case the physical parameters $\rho$, $\lambda$, $\rho_{p}$ and kinematical 
parameters $\theta$, $\sigma$, $\dot{u}_{i}$ and $V^{3}$ for the model (\ref{eq57}) 
are given by
\begin{equation}
\label{eq58}
\rho = \frac{1}{\mu(x)^{4n(2n + 1)}}\left[-\frac{8(2n + 1)d_{1}d_{2}r^{2}}{\mu(x)^{2}} + 
\frac{(2n + 1)^{2}(4n - 3)r^{2}\nu(x)^{2}}{\mu(x)} + \ell^{2}\right],
\end{equation}
\begin{equation}
\label{eq59}
\lambda = \frac{1}{\mu(x)^{4n(2n + 1)}}\left[-\frac{(4n + 1)(2n + 1)^{2}r^{2}\nu(x)^{2}}
{\mu(x)^{2}} + \ell^{2}\right],
\end{equation}
\begin{equation}
\label{eq60}
\rho_{p} =\frac{2}{\mu(x)^{4n(2n + 1)}}\left[-\frac{4(2n + 1)d_{1}d_{2}r^{2}}{\mu(x)^{2}} + 
\frac{(2n + 1)^{2}(4n - 1)r^{2}\nu(x)^{2}}{\mu(x)^{2}}\right],
\end{equation}
\begin{equation}
\label{eq61}
\theta = 0,
\end{equation}
\begin{equation}
\label{eq62}
\sigma^{2} = \frac{\ell^{2}}{\mu(x)^{4n(2n + 1)}},
\end{equation}
\begin{equation}
\label{eq63}
\dot{u}_{i} = \Bigl(\frac{2n(2n + 1)r \nu(x)}{\mu(x)}, 0, 0, 0 \Bigr),
\end{equation}
\begin{equation}
\label{eq64}
V^{3} = \sqrt{-g} = \mu(x)^{2(2n + 1)^{2}},
\end{equation}
where
$$\mu(x)= d_{1}e^{rx} + d_{2}e^{-rx}, $$
\begin{equation}
\label{eq64}
\nu(x) = d_{1}e^{rx} - d_{2}e^{-rx}. 
\end{equation}
\section{Concluding Remarks}
In the study, we have presented a new class of exact solutions of Einstein's field 
equations for inhomogeneous cylindrically symmetric space-time with string sources 
which are different from the other author's solutions. In these solutions 
all physical quantities depend on at most one space coordinate and the time.\\

In case (i), the models (\ref{eq39}) represents expanding, shearing and non-rotating 
universe. The expansion in the model increases as time increases when $n < 0$ but the 
expansion in the model decreases as time increases when $n > 0$. The spatial volume 
increases as time increases. If we set the suitable values of constants, we find that 
energy conditions $\rho \geq 0$, $\rho_{p} \geq 0$ are satisfied. We observe that 
$\frac{\sigma}{\theta}$ is constant throughout. Shear $(\sigma)$ vanishes when $p = 0$. 
Thus the model isotropizes when $p = 0$. The acceleration vector $\dot{u}$ is zero for 
$n = 0$. In this model all the physical and kinematic parameters vanish for $p = 0$, 
$q = 0$. The model (\ref{eq39}) represents a realistic model. In this solution 
all physical and kinematical quantities depend on at most one space coordinate and the 
time.\\

In case (ii), the expansion $\theta$, in model (\ref{eq57}), is zero. With the help of 
physical and kinematical parameters, we can determine some physical and geometric features 
of the model. All kinematical quantities are independent of $T$. In general, the model 
represents non-expanding, non-rotating and shearing universe. The spatial volume $V$ 
increases as distance increases. The shear $(\sigma)$ vanishes when $\ell = 0$. The 
acceleration vector $\dot{u}$ is zero for $n = 0$, $n = - \frac{1}{2}$. Choosing suitable 
values for constants $n$ and $L$, we find that energy conditions $\rho \geq 0$, $\rho_{p} 
\geq 0$ are satisfied. The solutions identically satisfy the Bianchi identities given by 
(\ref{eq19}) and (\ref{eq20}). In this solution all physical and kinematical quantities 
depend on at most one space coordinate.
\section*{Acknowledgements} 
One of the Authors (A. P.) would like to thank Professor G. Date, IMSc., Chennai, India 
for providing facility where part of this work was carried out. Authors also thank 
Professor Raj Bali for his fruitful suggestions which culminated the paper in present 
form. 

\end{document}